\documentclass[pra,reprint, amsmath,amssymb, aps,floatfix,]{revtex4-2}
\pdfoutput=1

\usepackage{graphicx}
\usepackage{subcaption}
\usepackage{dcolumn}
\usepackage{bm}
\usepackage{multirow}
\usepackage{array} 
\usepackage{makecell}
\usepackage[permil]{overpic}
\usepackage{orcidlink}
\usepackage{braket}
\usepackage{placeins}
\usepackage{float}
\usepackage{wrapfig}
\usepackage{caption}
\usepackage{float}
\usepackage{hyperref}

\begin{document}

\preprint{APS/123-QED}

\title{Quantum Convolutional Neural Networks for Jet Images Classification}

\author{Hala Elhag\, \orcidlink{0000-0003-1638-8022}}
\email{hala.elhag@desy.de}
\affiliation{Deutsches Elektronen-Synchrotron DESY, 15738 Zeuthen, Germany} 
\affiliation{ Institut für Physik, Humboldt-Universität zu Berlin, 12489 Berlin, Germany}

\author{Tobias Hartung}
\affiliation{Northeastern University - London, Devon House, St Katharine Docks, London, E1W 1LP, United Kingdom}
\affiliation{Khoury College of Computer Sciences, Northeastern University, \#202, West Village Residence Complex H, 440 Huntington Ave, Boston, MA 02115, USA}

\author{Karl Jansen}
\affiliation{Computation-Based Science and Technology Research Center, The Cyprus Institute, 20 Kavafi Street, 2121 Nicosia, Cyprus}
\affiliation{Deutsches Elektronen-Synchrotron DESY, 15738 Zeuthen, Germany}

\author{Lento Nagano}
\affiliation{International Center for Elementary Particle Physics (ICEPP), The University of Tokyo, 7-3-1 Hongo, Bunkyo-ku, Tokyo 113-0033, Japan}

\author{Giorgio Menicagli Pirina}
\affiliation{Northeastern University - London, Devon House, St Katharine Docks, London, E1W 1LP, United Kingdom}

\author{Alice Di Tucci}
\affiliation{Deutsches Elektronen-Synchrotron DESY, 15738 Zeuthen, Germany}
\affiliation{Physikalisch-Technische Bundesanstalt PTB, 38116 Braunschweig}

\date{\today}

\begin{abstract}
Recently, interest in quantum computing has significantly increased, driven by its potential advantages over classical techniques. Quantum machine learning (QML) exemplifies one of the important quantum computing applications that are expected to surpass classical machine learning in a wide range of instances. This paper addresses the performance of QML in the context of high-energy physics (HEP). As an example, we focus on the top-quark tagging, for which classical convolutional neural networks (CNNs) have been effective but fall short in accuracy when dealing with highly energetic jet images. In this paper, we use a quantum convolutional neural network (QCNN) for this task and compare its performance with CNN using a classical noiseless simulator. We compare various setups for the QCNN, varying the convolutional circuit, type of encoding, loss function, and batch sizes. For every quantum setup, we design a similar setup to the corresponding classical model for a fair comparison. Our results indicate that, using a classical simulator, QCNN with proper setups tend to perform better than their CNN counterparts, especially when the convolution block has a lower number of parameters. For the higher parameter regime, the QCNN circuit was adjusted according to the dimensional expressivity analysis (DEA) to lower the parameter count while preserving its optimal structure. The DEA circuit demonstrated improved results over the comparable classical CNN model.

\end{abstract}

\maketitle

\section{\label{sec:level1}Introduction}

Due to its potential advantages over classical techniques, the interest in quantum computing has lately rapidly increased. Quantum machine learning (QML)(e.g.~\cite{Schuld_2014,Biamonte_2017,Mangini_2021,challenges_cerezo}) serves as an important application of quantum computing that can potentially revolutionize data processing and analysis. 
In particular, we focus on quantum neural networks (QNN)~\cite{Mitarai_2018_paramShift,farhi2018classification,Benedetti:2019inj, PhysRevA.101.032308}
, where the task of quantum computing is to calculate a loss function with trainable parameters that are then optimized classically. However, to date, there is no established method to determine which type of data can be efficiently utilized in QML and how they have to be embedded into a quantum circuit to surpass classical machine learning.
Additionally, there is a lack of a systematic approach to constructing a specific QNN model suitable for a given problem. Therefore, it is important to seek problem-specific construction. We are addressing these issues in the context of experimental high-energy physics (HEP)~(e.g.~\cite{Guan:2020bdl,felser2021quantuminspiredmachinelearninghighenergy,Araz_2021,Araz_2022,Gianelle_2022,Belis_2024}), with a particular focus on classifying quantum chromodynamics (QCD) jet images for top quark jet tagging. This problem is highly relevant in the field of beyond the Standard Model physics, where some of the current theories, in supersymmetry for example, propose the existence of new particles with a very large mass. If such particles existed in a particle collider experiment, like in the proposed High Luminosity Large Hadron Collider (HL-LHC) at CERN, they would be unstable and will decay into very highly energetic particles. The heaviest known particle in the Standard Model is the top quark, making it a good candidate to probe new physics. The top quark itself is unstable and will decay into a b-quark and a $W$-boson. In the leptonic channel, the $W$ decays into a lepton and a neutrino. One of the ways particle physicists observe the results of such a process in a particle collider experiment is by visualizing the energy and momentum of the final state detected particles. This can be done through the formation of a jet image, which represents the base of a jet cone plotted in a histogram filled with a fraction of the energies of the detected final state particles. The detected particles would form subjets in the full jet image. In the case of the leptonic channel decay mode of the $W$-boson from a highly boosted top quark, the angle between the b-quark and the lepton would be very small which may result in having the lepton subjet be indistinguishable from and included in the b-quark subjet. The formed image might be similar to a QCD image, that is an image of a jet formed by the background processes of QCD interactions at the collider experiment, thus increasing the complexity of tagging top quark jet images.         

Classical convolutional neural networks (CNNs) have been widely and successfully employed for top-quark tagging~\cite{PhysRevD.105.042005}.
However, they struggle to provide the required accuracy when faced with the highly energetic and complex top-quark jet images. The complexity of an image is defined by how dense the information it holds and how diverse its visual features are~\cite{heaps1999similarity}. Therefore, many parameters will be needed to extract such features from the images which is further discussed in Section~\ref{sec:cnn}.  In~\cite{Kasieczka:2019dbj}, various modern machine learning models, including CNN as an image-based model, were benchmarked for the top-tagging task. These models demonstrated comparative performance, with accuracies ranging from 90\% to 94\%, but required intensive trainable parameter counts ranging from 1000 to 1.46 millions. The latter parameter count corresponds to the state-of-the-art CNN-based architecture named ResNeXt. This architecture introduces a balance between model depth and width for efficient feature learning, however, at the cost of increased computational cost and memory usage~\cite{noor2025surveystateoftheartdeeplearning}.
Aiming to overcome this challenge of increased parameter count, 
our main objective is to identify a practically suitable QML architecture to efficiently classify top-quark and QCD jet images. We do this by quantifying its accuracy
and comparing its performance with a classical machine learning method. Specifically, we focus on the quantum convolutional neural network (QCNN), first introduced by Cong et al.~\cite{cong}, which leverages shallow-depth quantum circuits, making it suitable for the noisy intermediate-scale quantum (NISQ) devices currently accessible. This circuit shallowness also provides robustness against barren plateau issues, ensuring the trainability of the model~\cite{Arthur}. While these advantages might make QCNN classically simulable, as highlighted in~\cite{bermejo2024quantumconvolutionalneuralnetworks}, the classical simulation of QCNN can still be regarded as a quantum-inspired machine learning architecture. In fact, several previous works have demonstrated improved accuracy of QCNN over conventional classical CNNs under certain conditions~\cite{Tak, Samuel}.

In this study, we utilize publicly available JetNet library datasets~\cite{jetnet}, specifically the TopTagging dataset which contains top and QCD jets. The top jets in this dataset are in the hadronic channel.
The current stage of this study is to have a proof of principle that our quantum neural network can be utilized in HEP analysis such as in top-tagging. Hence, this work's analysis will involve the top decay's hadronic channel rather than the leptonic channel, since it's the provided channel in the TopTagging dataset. 
However, from the jet images pre-processing we could already notice the similarity of top and QCD images which yields the required complexity to test our quantum and classical models. 
After forming the jet images, the principal component analysis (PCA) is applied to reduce the dimensionality of the formed images. The quantum model is implemented on a classical noiseless simulator (PennyLane~\cite{Bergholm:2018cyq}), while the classical counterpart is built using the TensorFlow framework~\cite{tensorflow2015-whitepaper}. Since we are working with a classical dataset, incorporating an appropriate encoding in our quantum circuit model is essential. So we compare various setups for the QCNN, varying the convolutional circuit, type of encoding, loss function, and batch sizes. To achieve an optimal QCNN structure, we conduct a dimensional expressivity analysis (DEA) to remove less significant gates from the circuit while preserving maximal expressivity. We design a similar setup for every quantum setup to the corresponding classical model and use the resulting performance as a reference.

The structure of this paper is as follows. In Section~\ref{sec:theo_bkg}, the concept of boosted top-quarks originating from a $W'$ boson is presented. Section \ref{SecII} describes the methodology of our study. In this section, we start by introducing the main components of a basic classical CNN model. We then explain how these concepts can be translated into the quantum domain to construct a QCNN, which is subsequently analyzed using DEA. After that, we discuss the formation and the pre-processing of the jet images. The results and QCNN to CNN comparisons are presented in section \ref{secIII}. Then, we conclude and mention future directions and plans in section \ref{secIV}.      

\section{High-Energy Physics Background} \label{sec:theo_bkg}

The Standard Model of particle physics successfully describes the fundamental particles and three of the four fundamental forces, excluding gravity, governing their interactions. Despite its success, it is incomplete, as it does not explain phenomena like gravity, dark matter, or neutrino mass. Extensions beyond the Standard Model (BSM) aim to address these gaps, and searches for BSM particles are among central objectives at the LHC experiments. 

A leading candidate among the hypothetical particles proposed by various BSM theories is the $W'$ boson. This particle is considered a heavier analog of the Standard Model $W$ boson and is theorized to interact predominantly with third-generation fermions~\cite{thomson}, such as the top quark. A specific type, the right-handed $W'$ boson, is predicted within the left-right symmetric model, which is an extension of the Standard Model’s electroweak sector that unifies the gauge groups as $SU(2)_R \times SU(2)_L \times U(1)$ \cite{d0paper}.

The D0 Collaboration initiated one of the earliest searches for the $W'$ boson in 1995, focusing on its decay into an electron and an anti-neutrino. This study established exclusion limits on the $W'$ boson mass to 800GeV \cite{d0paper, muller}. Subsequent analyses, such as a 2017 study by the CMS Collaboration, placed a 95\% confidence level lower bound of 2.4TeV on the mass of a right-handed $W'$ boson decaying into a top and a bottom quark \cite{AN}.

\begin{figure}
    \centering
    \includegraphics[width=0.6\linewidth]{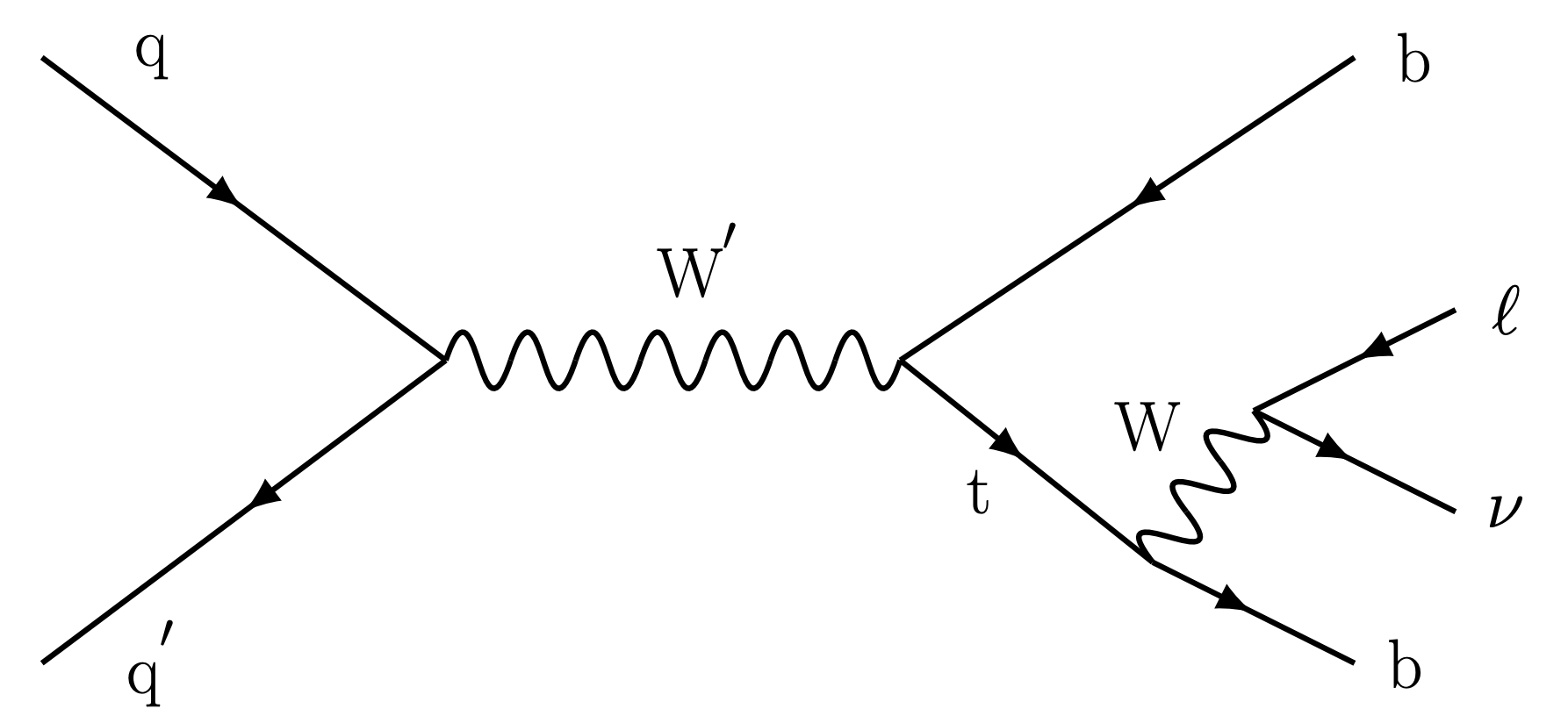}
    \caption{Leptonic decay channel of $W'$ boson to a top and a bottom quark ($W' \rightarrow t b$).}
    \label{fig:Wprime_decay}
\end{figure}

Because of its large predicted mass, the $W'$ boson is expected to decay promptly into other particles. One of the most studied decay channels is $W' \rightarrow t b$, which is especially compelling due to its relatively low background from QCD processes \cite{AN}. Figure~\ref{fig:Wprime_decay} depicts the leptonic decay of the $W'$ via this mode. Thus, developing an efficient and reliable approach for tagging boosted top-quark jets could provide deeper insights into the properties of the $W'$ boson in case it was produced in a particle collider.

Top quark identification in collider experiments involves detecting characteristic features such as secondary vertices from $b$-quark decays~\cite{thomson} and localized energy deposits in detectors corresponding to jets formed by the final-state particles. The neutrino, which escapes direct detection, is inferred from the missing transverse energy.

\begin{figure}[t]
    \centering
    \includegraphics[width=0.8\linewidth]{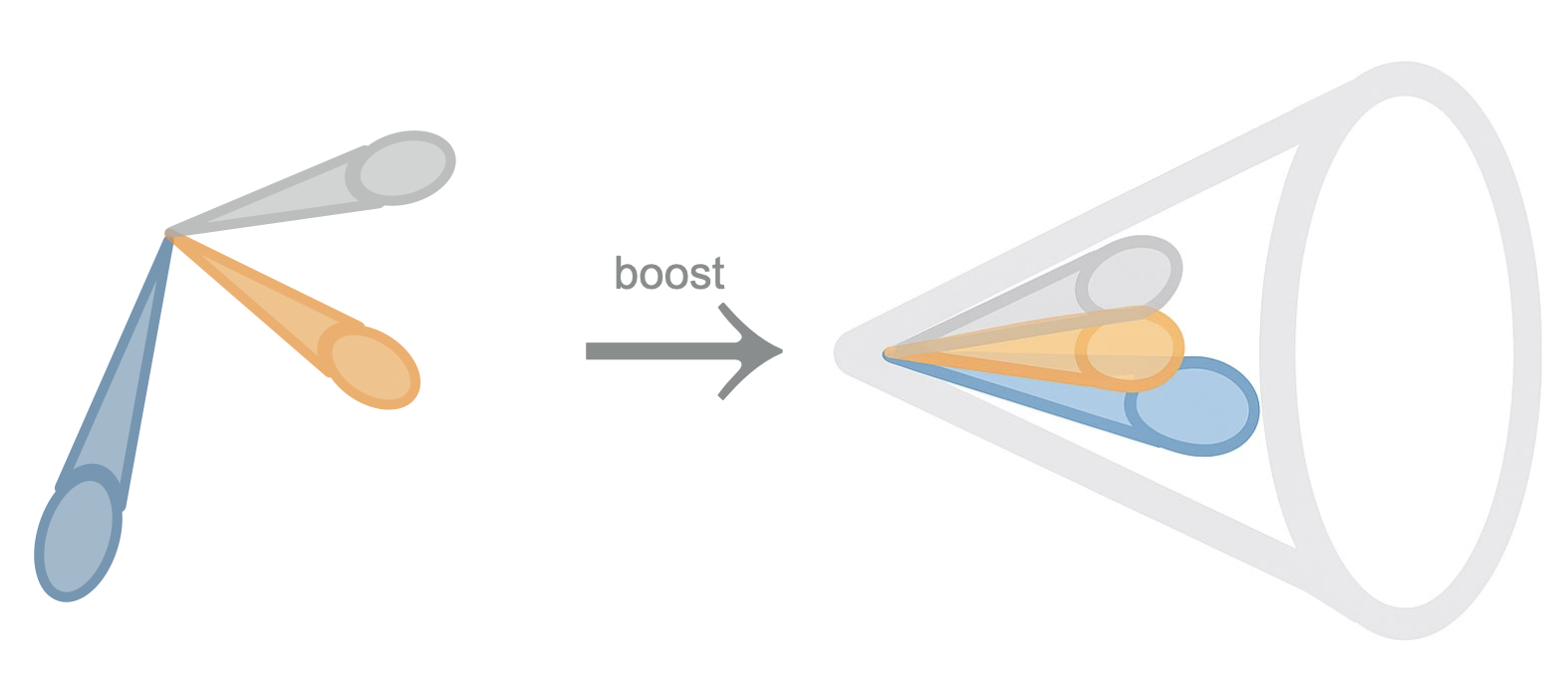}
    \caption{Illustration of boosted jets, e.g. decayed from a top quark.}
    \label{fig:boosted_top}
\end{figure}

Several methods are employed to analyze these decays. One approach involves creating a two-dimensional histogram of the energy distribution of decay products, typically a lepton and $b$-quark hadrons, which are represented as two subjets in the resulting jet image. The formation of these jet images is described in detail in Section~\ref{sec:jet_imgs_form}. The angular separation between the subjets is determined by the transverse momentum ($p_T$) of the top quark. When the top quark is highly boosted, this angle becomes small, leading to an overlap between the lepton and $b$-quark subjets. This phenomenon, illustrated in Figure~\ref{fig:boosted_top}, shows how boosted top jets, with high $p_T$, have closely aligned subjets, while non-boosted (low $p_T$) tops exhibit well-separated ones. This overlap can cause boosted top jet images to resemble those from QCD processes.

\section{Methodology}\label{SecII}

\subsection{\label{sec:cnn}CNN}

The CNN (see e.g.~\cite{2015arXiv151108458O} for a review) showed a wide range of implications in computer science as well as other fields of science, which makes it occupy an important position in machine learning, especially in image classification tasks. The basic idea behind CNN is suggested by its name, it mainly uses some filters, which convolve over an input (usually an image) to detect some patterns~\cite{tensorflow2015-whitepaper}. 
These patterns are usually referred to as features. If one aims to detect a specific object from an image, one may use feature mapping to spot the object feature and vague the other features. This is done by scanning each pixel in the image while checking the surrounding pixels. Then, multiply the values of these pixels by the corresponding weights of the used filter.

\begin{figure}[h]
    \centering
    \includegraphics[width=0.4\textwidth]{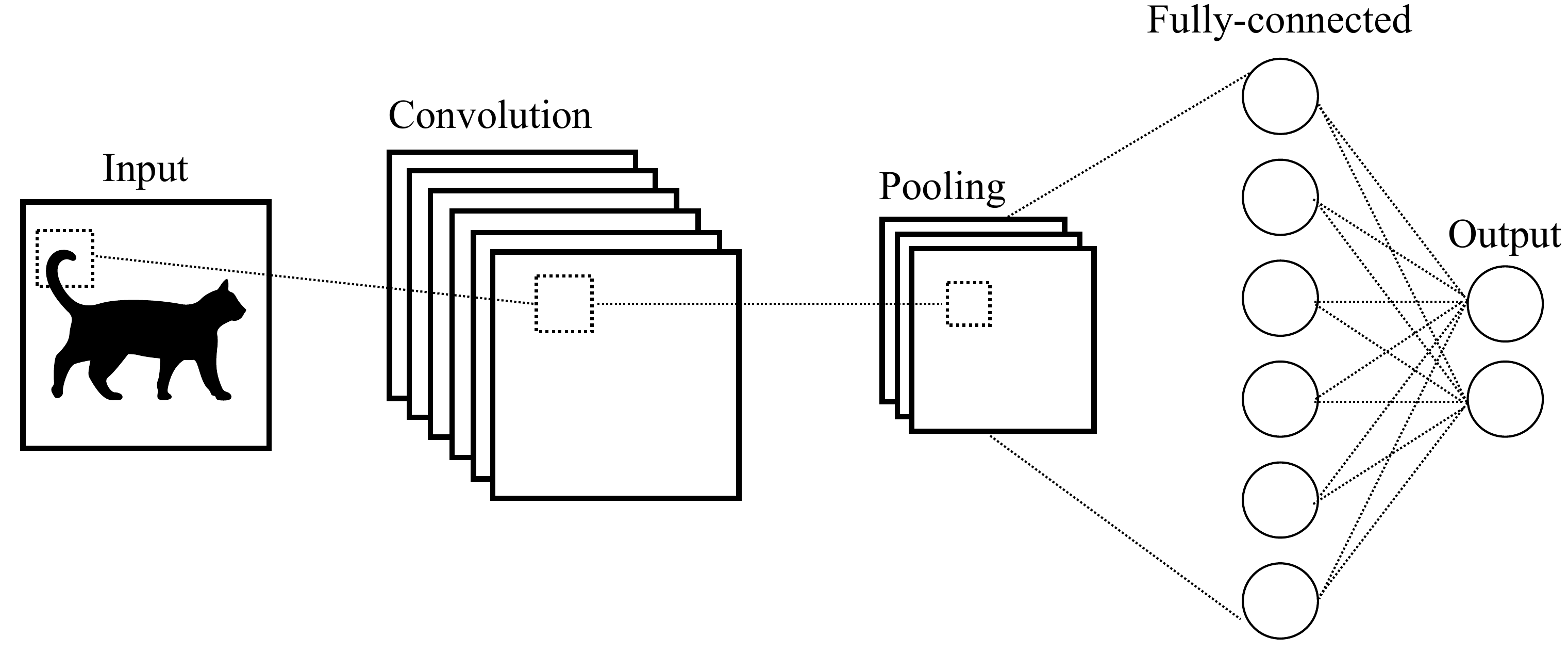}
    \caption{General structure of a CNN model. } 
    \label{FigCNN}
\end{figure}

There are usually several different layers associated with the CNN architectures. The maximum pooling (MaxPooling) layer comes after the convolutional layer (Conv2d). This MaxPooling layer has the 2D shaped window that takes all the neurons of the input and passes over the maximum pixel value to the next layer. So, it is used to reduce the dimensionality of the output of the previous convolutional layer while keeping the important features. Several of these layers can be applied before adding the flatten and the dense layers. The flatten converts the output of the previous layers from a multi-dimensional into a vector, which is the required input shape for the dense layer. Then, this output goes into the dense layer as an input, and then it gets connected to all the neurons of the previous layers to work as a classifier for the images. 
Finally, activation functions are usually added to the dense layers. The Rectified Linear Unit (ReLU) returns the same value of the input if it is positive, otherwise, it returns a zero. Other examples of activation functions are Sigmoid which outputs values between 0 and 1, and Tanh which outputs a value between -1 and 1. The last dense layer in the network represents the output layer, and it contains a number of neurons that is equal to the number of different classes we had for our inputted images. The value of each neuron indicates the probability that this input corresponds to a specific classification. Sigmoid or Tanh could also be used as an activation function in that dense layer, depending on the problem and the loss function used. In principle, passing through several convolutional layers makes the information more focused and accurate before it comes to the dense layer.
The general structure of CNN architectures can be viewed in Fig.~\ref{FigCNN}. In this project, we use only one Conv2d layer, followed by a MaxPooling layer, a flatten layer, and two dense layers at the end.

\begin{figure}[h]
    \centering
    \includegraphics[width=0.5\textwidth]{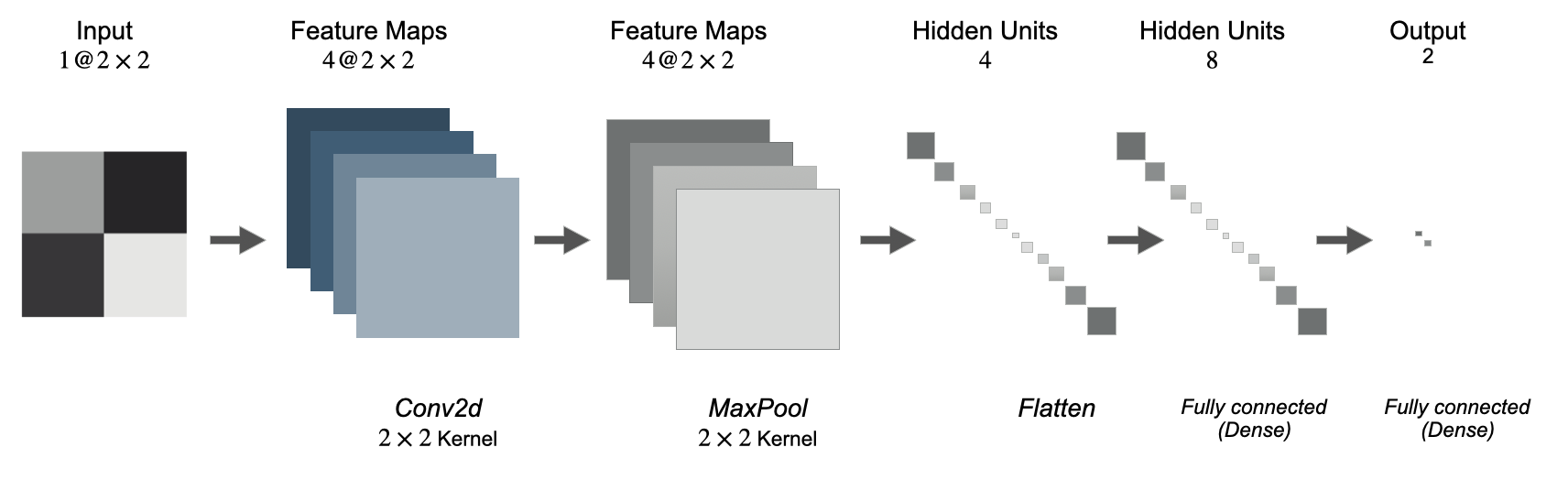}
    \caption{Detailed structure of the CNN model used in this study. } 
    \label{fig:CNN_struc}
\end{figure}

A detailed structure of the CNN model used in this study, consisting of 33 trainable parameters, is presented in Fig.~\ref{fig:CNN_struc}. The first layer, labeled Conv2d, contains four feature map kernels, each designed to extract specific patterns from the input image. Generally, as the complexity of input images increases, a larger number of feature maps are required to capture more features, which in turn leads to a higher number of trainable parameters. In other words, complex datasets require complex models that are capable of learning complex connections and patterns found in data points. This can significantly increase the computational cost, potentially making the model impractical for large or complex datasets. Such models also suffer from the overfitting issue which results from the increased number of trainable parameters which perfectly fits to the given dataset, but cannot generalize to new dataset~\cite{devshatwar2023challenges}. This challenge represents a key limitation of conventional CNNs and traditional machine learning models. For example, as previously mentioned in the introduction, ResNeXt representing a state-of-the-art CNN architecture involves number of parameters exceeding a million. Methods such as regularization and principal component analysis (PCA) are usually employed to decrease the complexity and increase generalization of machine learning models. The used case of this project utilizes PCA to reduce the dimensionality of the dataset which reduces the number of parameters. However, this might cause a reduction in the efficiency of the classical machine learning model.   

One of the goals of QML models is to address this issue by employing quantum entanglement, which allows the model to learn correlations between features across the image rather than relying on local information only. The main subject of this paper, QCNNs, are an example of such models and are discussed in Section~\ref{sec:qcnns}.

\subsection{Loss functions}
\label{subsec:loss-function}
The main mission one needs to accomplish when training the neural network is to minimize the difference between the output from the neural network and the true value as much as possible. 
The prediction is quantified by a metric called a loss function, which is optimized by adjusting the trainable parameters.
For training our CNN and QCNN models of this study, several loss functions listed below are tested.
 
1) The cross-entropy loss is a measure of the classifier performance with an output probability between 0 and 1, and given by
\begin{equation}
 \mathcal{L}_\text{Cross-Entropy} = -\frac{1}{N} \sum_{i=1}^{N} \left( y_i \log(\hat{y}_i) + (1 - y_i) \log(1 - \hat{y}_i) \right)\,,
\end{equation}
where $y_i$ is the true label of the $i$-th input, $\hat{y}_i$ is the corresponding label predicted by the model, and $N$ is the number of training data.

2) The Mean-Square-Error (MSE) considers binary labels~$\pm1$ and is given by 
\begin{equation}
    \mathcal{L}_\text{MSE} = \frac{1}{N} \sum_{i=1}^{N} (y_i - \hat{y}_i)^2\,.
\end{equation}

3) The Hinge loss also takes $\pm 1$ labels as inputs
and defined as
\begin{equation}
    \mathcal{L}_\text{Hinge} = \frac{1}{N} \sum_{i=1}^{N} \max(0, 1 - y_i \cdot \hat{y}_i)\,.
\end{equation}

Finally, these loss functions are optimized via the Adam optimizer~\cite{2014arXiv1412.6980K} in this work.
This method combines several algorithms such as momentum and RMSProp optimizers to obtain reliable learning results.

Due to GPU's limited capabilities, it is not possible to train the network with the whole dataset all at once when the dataset is too large. Therefore, the data is usually split into batches, where the number of samples that go through the network defines the batch size. When the network goes through the whole dataset, this is one epoch.
Then this process is repeated until the loss function converges during optimization.

\subsection{QCNNs}\label{sec:qcnns}

\begin{figure*}
\centering
\includegraphics[width=120mm]{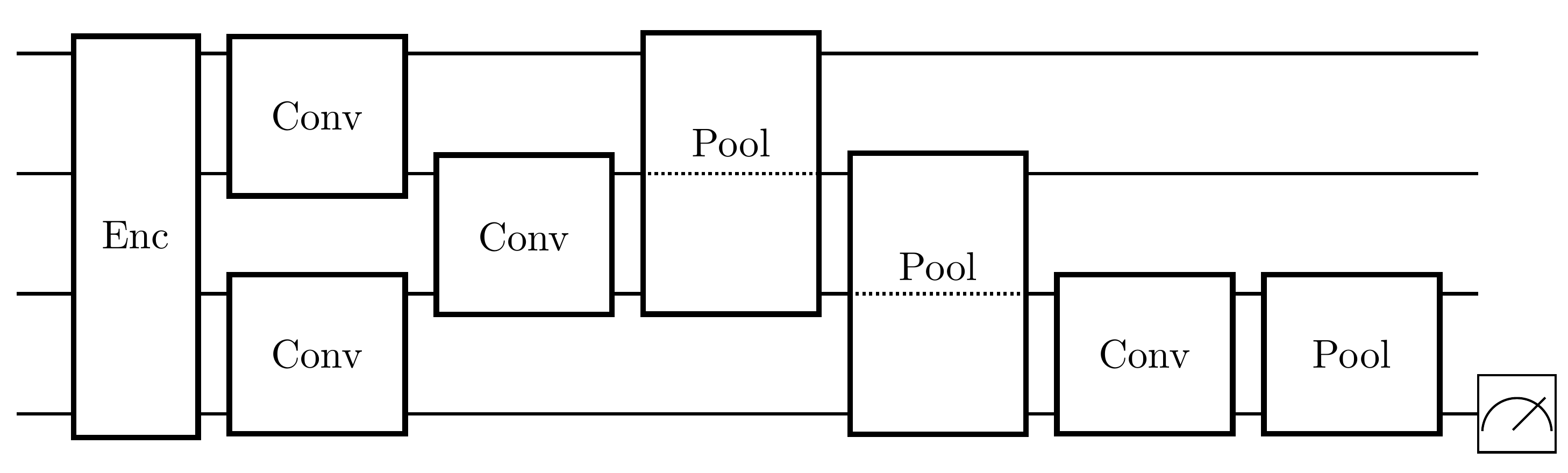}
\caption{The structure of a full QCNN circuit. The first block (Enc.) corresponds to one of encodings given in Fig.~\ref{FigEnc}, and the following blocks (conv. and pool.) are given in Fig.~\ref{FigQCP}} 
\label{FigQCNN}
\end{figure*}

QCNN was a QNN model motivated by the success of CNNs~\cite{cong}. The basic idea of QCNN is to utilize quantum operations to accomplish the convolution and pooling techniques and do the classification task using the outcome from the circuit. 
In the case of classical inputs (like images), we need to encode the data (image pixels) first 
as $\ket{\psi(\bm{x})}=U_{\text{enc}}(\bm{x})\ket{0}$, where $\bm{x}$ is classical data. 
Then the encoded state is processed via parametrized unitaries composed of convolutional and pooling blocks to have an output $\hat{y}$, given by
\begin{equation}
    \hat{y}=\braket{\psi(\bm{x})|U^{\dag}_{\text{QCNN}}(\bm{\theta}) Z_{N_q}U_{\text{QCNN}}(\bm{\theta})|\psi(\bm{x})}\,,
\end{equation}
where $N_q$ is the number of qubits.
The overall structure for the QCNN with classical inputs is shown in Fig~\ref{FigQCNN}.

\subsubsection{Encodings}
\begin{figure}[ht]
    \centering
    \includegraphics[width=85mm]{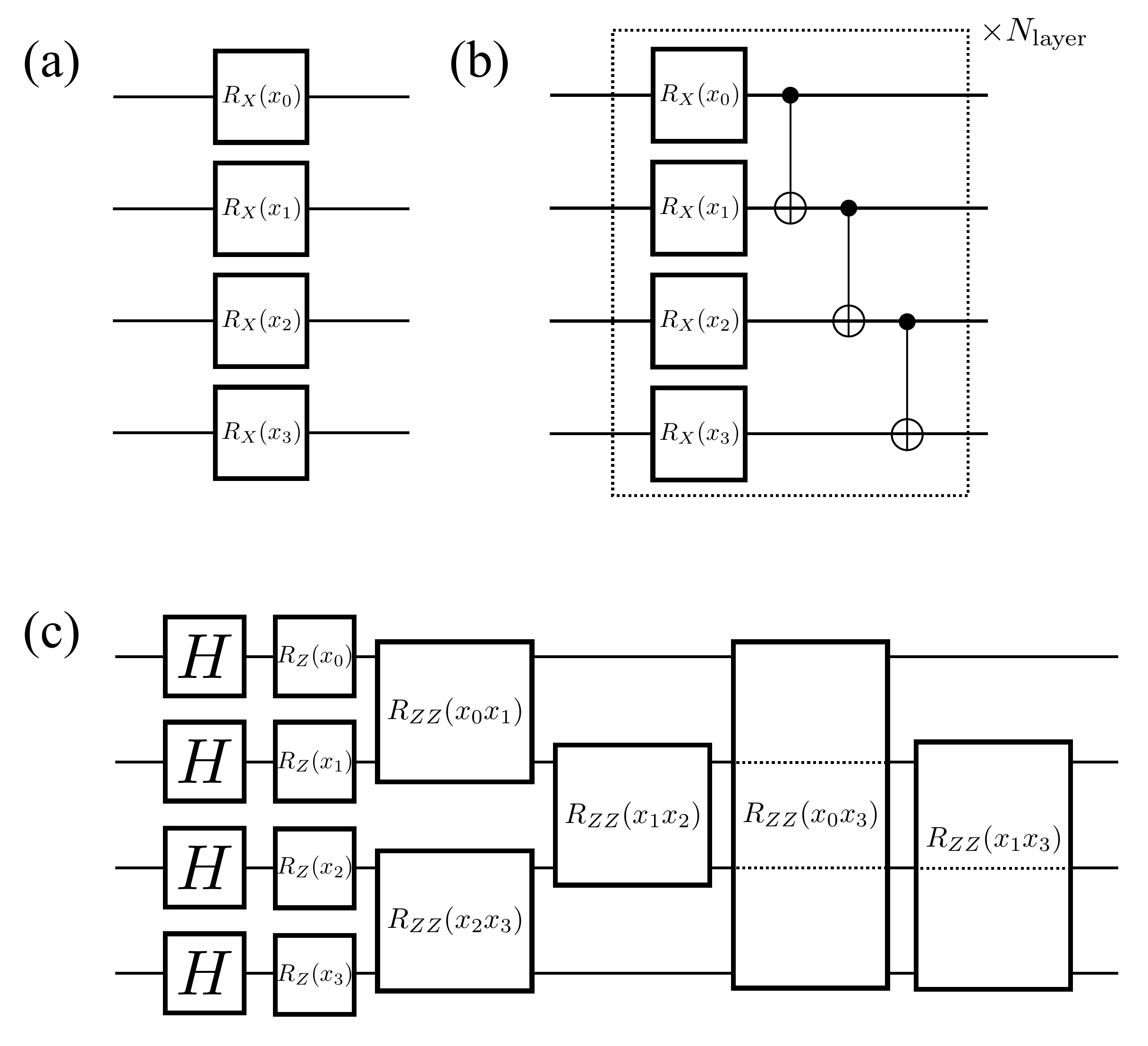}
    \caption{The four different encoding studied in this paper. TPE is represented in (a), HEE ($N_{\text{layer}}=1$ for HEE1 and $N_{\text{layer}}=2$ for HEE2) (b), and CHE (c).} 
    \label{FigEnc}
\end{figure}

The encoding $V_{\text{enc}}$ can be done in several ways. We test four encoding schemes, namely: tensor product embedding (TPE) which in this case is also known as angle encoding, hardware efficient embedding (HEE), and Classically Hard Embedding (CHE). These three encodings are used to study the effects of encoding on trainability in~\cite{2021arXiv211014753T}. 
The quantum circuit structure of the different encodings can be seen in Fig.~\ref{FigEnc}, where TPE is shown in Fig.~\ref{FigEnc}a, HEE in \ref{FigEnc}b, and CHE in~\ref{FigEnc}c. 
The data inputs in these figures are represented by $x_i$, where $i \in\{0,1,2,3\}$. In this paper, we use one layer and two layers of HEE, denoted by HEE1 and HEE2, respectively.
On the other hand,
only one layer of CHE is used in this work.
Since reference~\cite{2021arXiv211014753T} showed that increasing the number of layers would cause trainability issues, we only focus on these encodings with the limited number of layers.

\subsubsection{Convolutional and Pooling circuits}

\begin{figure}[ht]
    \centering
    \includegraphics[width=80mm]{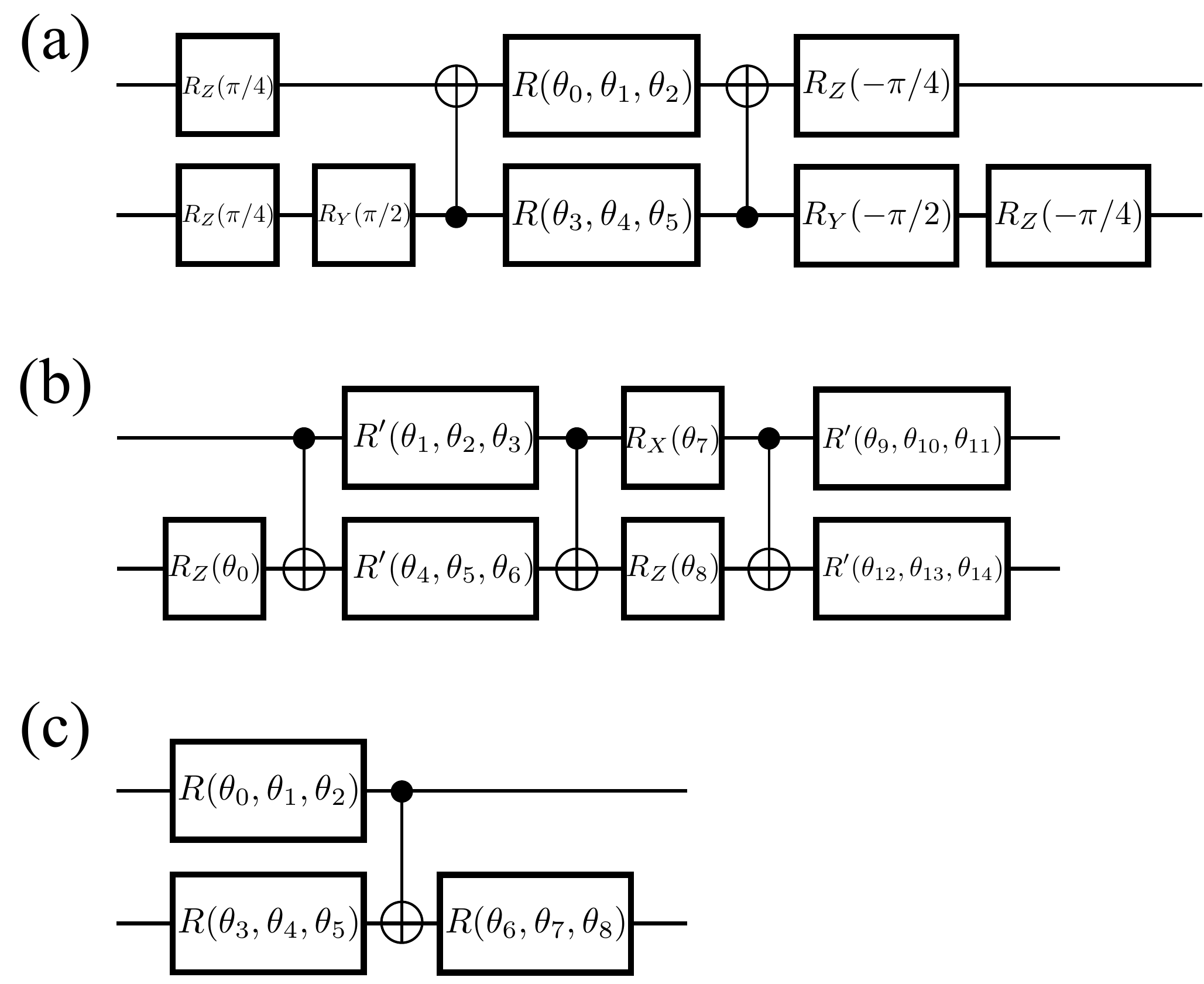}
    \caption{The quantum gates structure of $SO(4)$ (a) and $SU(4)$ (b) convolutional circuits, and pooling circuit (c), where $R(\alpha,\beta,\gamma)=R_Z(\gamma)R_Y(\beta)R_Z(\alpha)$ and $R'(\alpha,\beta,\gamma) = R_Z(\gamma)R_X(\beta)R_Z(\alpha)$.} 
    \label{FigQCP}
\end{figure}

After the encoding layer comes the convolutional layer. This layer consists of convolutional circuits connecting every pair of qubits to compensate for the task of a classical convolutional filter that scans over all the pixel values of an image. The convolutional circuit can be built in various ways. In this project, we test two unitary two-qubit circuits $SO(4)$ and $SU(4)$.
The dimensionality of the trainable parameters $\bm{{\theta}}$ depends on these two-qubit unitaries. 
The circuit constructions for these unitaries are presented in Fig.~\ref{FigQCP}a and~\ref{FigQCP}b. The $SO(4)$ circuit corresponds to real transformation~\cite{farrokh} and it contains 2 CNOTs and 12 single-qubit gates of which only 6 of them contain trainable parameters ($\bm{{\theta}}$). $SU(4)$ is more general and considered a universal two-qubit circuit spanning the whole Hilbert space. It consists of 3 CNOTs and 15 single-qubit gates all containing trainable parameters~\cite{farrokh,PhysRevA.69.062321}.
Then, the convolutional circuit connects every pair of neighboring qubits until it forms a full convolutional layer. 

After that comes the pooling layer to reduce the dimensionality of the system by half. The pooling circuit, shown in Fig.~\ref{FigQCP}c, comprises a CNOT gate and 9 single-qubit gates all containing trainable parameters. The convolutional and pooling layers are repeated until the system has reached a sufficiently small size, one qubit in our case.  The output of this qubit is measured with a PauliZ measurement. The prediction of the neural network corresponds to the expectation value of the PauliZ operator and is then compared to the true label for the classification task as explained~in~Section~\ref{subsec:loss-function}.

\subsection{DEA}

Dimensional expressivity analysis (DEA), introduced by Funcke et al. in \cite{funcke2021dimensionalexpressivityanalysisbestapproximation} and \cite{Funcke}, is a technique designed to optimize a parametric quantum circuit (PQC) by achieving maximal expressivity with the fewest trainable parameters. In other words, it enables the construction of a PQC that retains its full expressive capability while minimizing redundancy in its parameters. Maximal expressivity here refers to the ability of the circuit to generate all relevant physical states, while a minimal set of parameters ensures that only independent trainable parameters remain, with any redundant ones either eliminated or set constant.

To formalize this, let us consider the PQC as a map~$C$ that takes the parameters from the parameter space~$\mathcal{P}$ to the state space $\mathcal{S}$. For any $\theta \ \epsilon \  \mathcal{P}$, the output $C(\theta)$ would be a state generated by the quantum circuit.

To identify redundant parameters, we first check the derivative of the PQC map with respect to each parameter $\theta_k$, denoted as $\partial_k C(\theta)$. If $\theta_k$ is redundant, then $\partial_k C(\theta)$ can be expressed as a linear combination of the remaining partial derivatives $\partial_j C(\theta)$ for $j \neq k$. A clear way to see this is through the Jacobian $J$ of $C$ 

\begin{equation}
    J_k(\theta) = \begin{pmatrix} \mathcal{R}\partial_1 C(\theta) &... & \mathcal{R}\partial_k C(\theta) \\ \mathcal{I}\partial_1 C(\theta) &... & \mathcal{I}\partial_k C(\theta)\end{pmatrix},
\end{equation}
where we check the rank of the $J$ inductively for each~$k$. If adding a column $\partial_k C(\theta)$ does not increase the rank of~$J$, then $\theta_k$ is deemed redundant and can be eliminated. Once the rank of~$J$ found equal the dimension of $\mathcal{S}$, all remaining parameters are redundant and no further checks are necessary.

Another efficient way to identify redundant parameters is through the $S$-matrix, which is defined as 
\begin{equation}
    S_k = J_k^* J_k.
\end{equation}
The matrix $J_k$ has a full rank, if and only if all eigenvalues of $S_k$ are positive. 
If, however, the smallest eigenvalue of $S_k$ was found to be zero, then we know that $\theta_k$ is redundant.

\subsection{Dataset Preparation}

As previously mentioned, the publicly available TopTagging dataset from the JetNet library~\cite{jetnet} is used. The original given dataset consists of two types of jets: top jets and QCD jets. The data is given in the form of a group of particles in every jet, where every particle has its own four-momentum components. The general form of the four-momentum vector is given by
\begin{equation}
    p^\mu = (E, p_x, p_y, p_z)\,,
\end{equation}
where its components are referred to as the particle's features. The maximum number of particles provided in each jet is 200. The dataset is then preprocessed and transformed to form images. Then, we apply PCA to reduce the dimensionality of the images. This reduction enables the use of a simple quantum circuit where each qubit takes a single pixel value from the image as an input.

\subsubsection{JetNet Dataset}

JetNet is a Python library developed to enhance accessibility and reproducibility in machine learning research within HEP, with a main focus on particle jet analysis. It is built on PyTorch framework and provides interfaces for accessing a range of HEP datasets, along with built-in evaluation metrics and general-purpose tools tailored for HEP tasks~\cite{jetnet}.

The library includes three main dataset categories: JetNet, QuarkGluon, and TopTagging. The JetNet dataset contains jets initiated by gluons, top quarks, light quarks, as well as W and Z bosons. As its name suggests, the QuarkGluon dataset consists of jets originating from quarks and gluons. The TopTagging dataset, which is used in the analysis presented in this study, features hadronic top jets as signal and QCD jets as background. All datasets are based on Monte Carlo simulations, which model particle collisions and detector responses to replicate real experimental conditions~\cite{AtlasMC}.

\begin{figure}
    \centering
    \includegraphics[width=\linewidth]{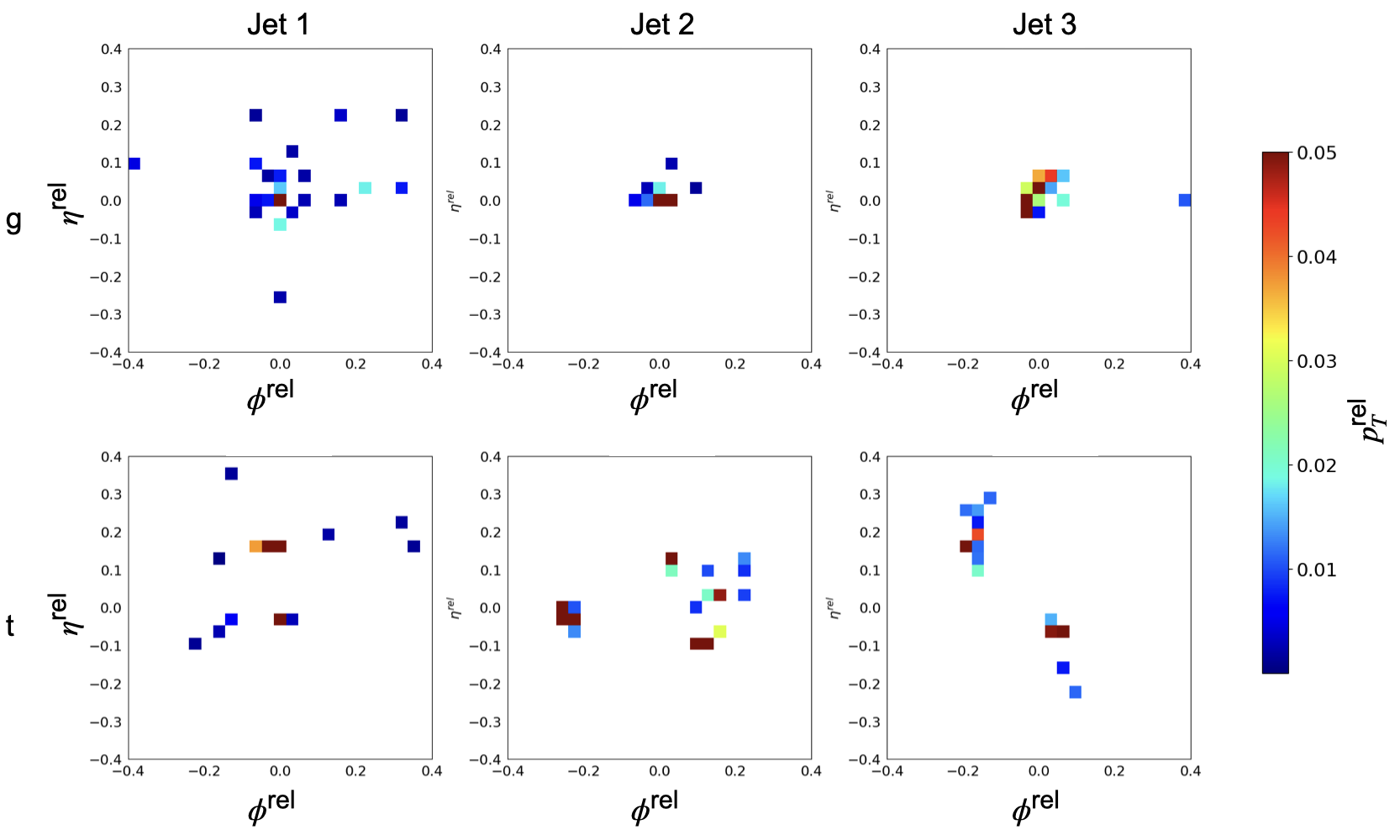}
    \caption{Jet images of a single-jet each of gluon (g) and top (t) particles~\cite{jetnet}.}
    \label{fig:jetnet_images}
\end{figure}

In addition, the JetNet dataset offers functionality for visualizing jet images using histogram plots. Some of these examples are shown in Fig.~\ref{fig:jetnet_images}. Each jet image corresponds to an individual jet from a separate event, mapped onto the $\eta$–$\phi$ plane. The intensity of each pixel reflects the transverse momentum of the jet’s constituents at that location. These individual images usually do not reveal any clear structure or consistent pattern within a given class. To uncover broader features, particle physicists often sum and normalize thousands of such jet images, creating an average or composite image that emphasizes recurring patterns not visible in single-event displays, which will be presented in the next section.

\subsubsection{Jet images}\label{sec:jet_imgs_form}

\begin{figure*}[hbt!]
    \centering
    \includegraphics[width=0.8\textwidth]{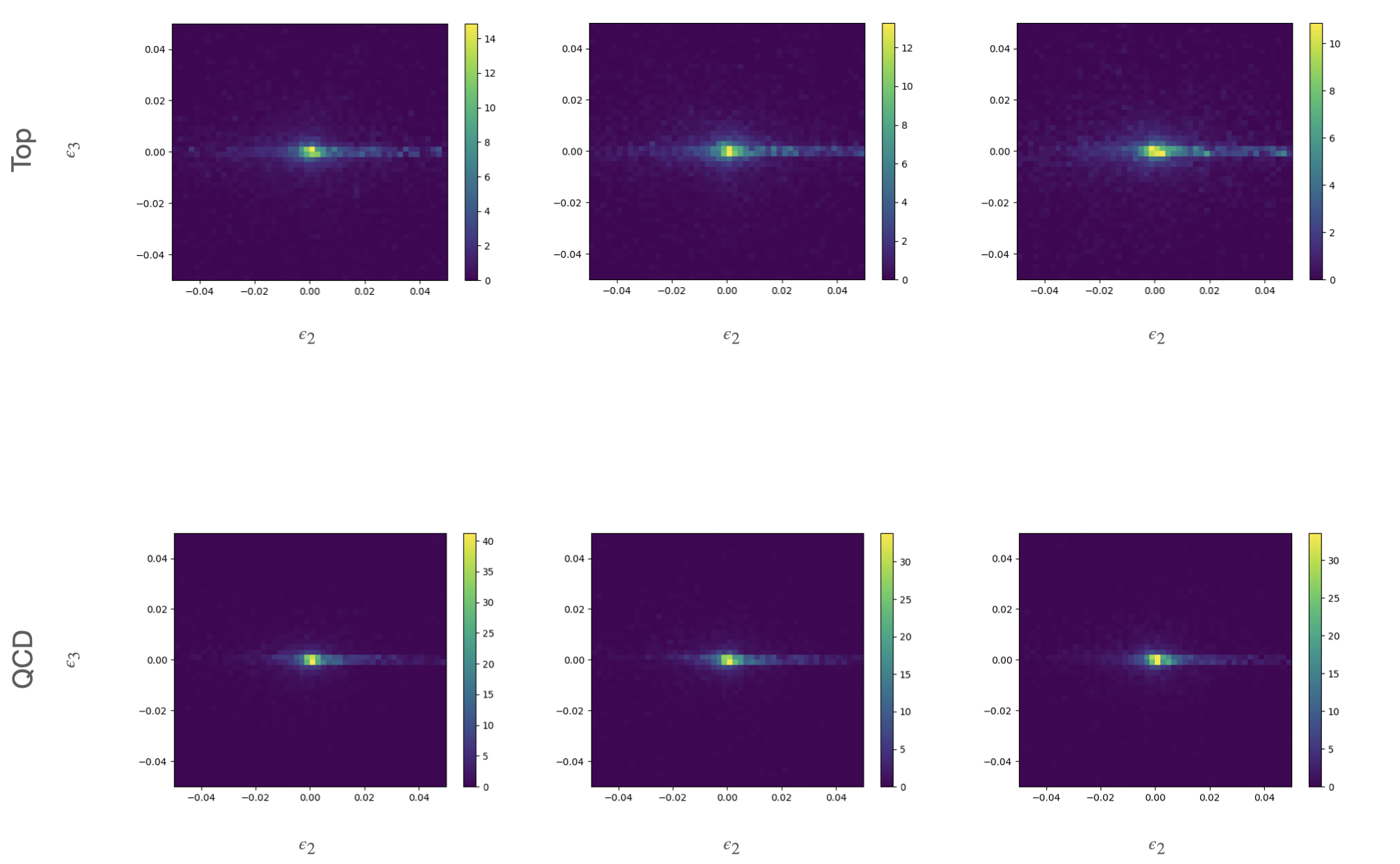}
    \caption{Examples of 50x50 jet images of top-quark (top) and QCD (bottom) jets from JetNet library.} 
    \label{FigJets}
\end{figure*}

Pre-processing and transformation techniques used to form the jet images are presented in references~\cite{PhysRevD.105.042005} and~\cite{2019arXiv190302032R}.
They have resulted in a highly reduced dependence for the jet images on the initial top quark momentum. 

The idea of image processing is as follows. First, we re-scale the jet four-momentum such that its mass is $m_B$. Then, we apply a Lorentz boost such that the jet energy is $E_B$ in the new frame. Using $m_B$ and $E_B$ a fixed value for the Lorentz boost can be obtained from
\begin{equation}
    \gamma_B = \frac{E_B}{m_B}\,.
\end{equation}
In this project, $\gamma_B=10$ is chosen which produced a convenient jet shape compared to several values arbitrarily tested.

The transformation of the jet's constituents is done using the Gram-Schmidt method. First, we choose the three constituents with the highest three-momentum ($\vec{\textbf{p}}_1$, $\vec{\textbf{p}}_2$, and $\vec{\textbf{p}}_3$) such that the total momentum of the jet is roughly 
\begin{equation}
    \vec{\textbf{P}}_J = \vec{\textbf{p}}_1 +\vec{\textbf{p}}_2 + \vec{\textbf{p}}_3.
\end{equation}
Three vectors~$\vec{\textbf{P}}_J, \vec{\textbf{p}}_1$, and $\vec{\textbf{p}}_2$ can be considered as a set of linearly independent vectors, hence we can use them to construct the Gram-Schmidt orthonormal set of basis. We may choose one of the vectors in the set $\{ \vec{\textbf{P}}_J, \vec{\textbf{p}}_1, \vec{\textbf{p}}_2 \}$ to form the first Gram-Schmidt basis, and let it be $\vec{\textbf{P}}_J$ so that 
\begin{equation}
    \hat{\epsilon}_1 = \frac{\vec{\textbf{P}}_J}{|\vec{\textbf{P}}_J|}\,.
\end{equation}

The second basis has to be orthogonal to $\hat{\epsilon}_1$. Therefore, we can use the vector $\vec{\textbf{p}}_1$ and subtract from it the component that projects on $\hat{\epsilon}_1$ such that 

\begin{equation}
    \hat{\epsilon}_2 = \frac{\vec{\textbf{p}}_1 - (\vec{\textbf{p}}_1 \cdot \hat{\epsilon}_1)\hat{\epsilon}_1}{|\vec{\textbf{p}}_1 - (\vec{\textbf{p}}_1 \cdot \hat{\epsilon}_1)\hat{\epsilon}_1|}\,.
\end{equation}
Then the third basis that is orthogonal to $\hat{\epsilon}_1$ and $\hat{\epsilon}_2$ reads
\begin{equation}
    \hat{\epsilon}_3 = \frac{\vec{\textbf{p}}_2 - (\vec{\textbf{p}}_2 \cdot \hat{\epsilon}_1)\hat{\epsilon}_1 - (\vec{\textbf{p}}_2 \cdot \hat{\epsilon}_2)\hat{\epsilon}_2}{|\vec{\textbf{p}}_2 - (\vec{\textbf{p}}_2 \cdot \hat{\epsilon}_1)\hat{\epsilon}_1 - (\vec{\textbf{p}}_2 \cdot \hat{\epsilon}_2)\hat{\epsilon}_2|}\,.
\end{equation}

The jet axis of the image is parallel to the first basis $\hat{\epsilon}_1$, hence the two-dimensional coordinates of the jet images $(X_i,Y_i)$ are related to the jet constituent $i$ having the four-momentum $p^\mu_i=(p^0_i,\vec{p}_i)$ through the formulae 
\begin{equation}
    X_i = \frac{\vec{p}_i \cdot \hat{\epsilon}_2}{p^0_i} \,,
\quad
    Y_i = \frac{\vec{p}_i \cdot \hat{\epsilon}_3}{p^0_i}\,.
\end{equation}

If we neglect the mass of the constituent, and as a consequence of the Gram-Schmidt unitary transformation that preserves the magnitudes of the transformed vectors, we would have 
\begin{equation}
    |\vec{p}_i|^2 = (p^0_i)^2 = (\vec{p}_i \cdot \hat{\epsilon}_1)^2 + (\vec{p}_i \cdot \hat{\epsilon}_2)^2 + (\vec{p}_i \cdot \hat{\epsilon}_3)^2.  
\end{equation}
This means that the components $X_i$ and $Y_i$ can only be in the range of $[-1,1]$. These components then form the coordinates of the two-dimensional histogram corresponding to the jet image. The histogram is filled with the weight 
\begin{equation}
    \omega_i = \frac{p^0_i}{E_B}, 
\end{equation}
representing the fraction of energy for the constituent~$i$. The goal of the Gram-Schmidt transformation is to make the jet images look similar regardless of the difference in the initial top-quark boost. This helps the machine learning networks to learn the features from images more easily. Examples of top-quark and QCD jets summed over 10,000 images are shown in Fig. \ref{FigJets}. We can already notice the similarity between the top and QCD jet images which increases the complexity of the analysis. The main differences one could notice are the slightly larger spread of the top jet compared to the QCD jet and the values of the energy fractions given in the color bars beside each histogram. We use 2,000 of these images, resized to~$28\times28$ pixels, for our training task in QCNN and CNN. The dataset is split into 80$\%$ for training and 20$\%$ for testing.

\subsubsection{PCA} \label{2.3}

Principal Component Analysis (PCA) (see e.g.~\cite{jolliffe2016principal})
is a method that is commonly used to find patterns while reducing the dimensionality of data. It mainly captures correlation in data and transforms it into a new set of variables that better describes the patterns. These variables are called the principal components. 

\begin{figure*}[hbt!]
    \centering
    \includegraphics[width=0.9\textwidth]{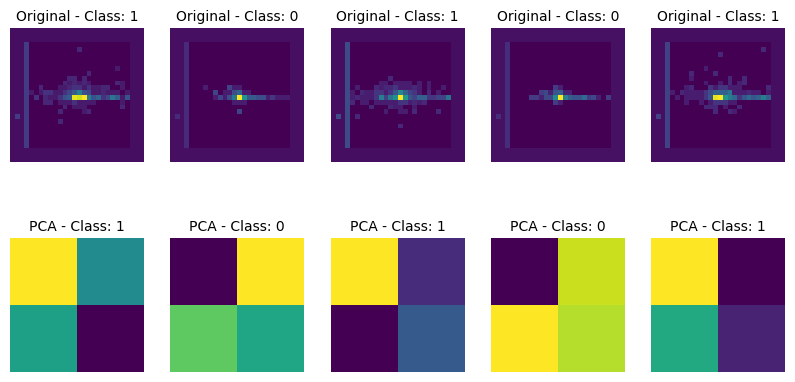}
    \caption{Jet images of top-quark (Class 1) and QCD (Class 0) originally with size 28x28 (top) compared to the PCA reduced of 2x2 (bottom).} 
    \label{fig:pca_jetImg}
\end{figure*}

In this study, PCA is used to reduce the dimensionality of the jet images from $28\times28$ to $2\times2$, so that we end up with only 4 pixels that are the inputs of our QCNN and CNN models. This implies that the original pixel data are projected onto the four principal eigenvectors of the data's covariance matrix, which are those associated with the largest eigenvalues. These principal components preserve most of the essential information from the original images, providing a compact and efficient representation of the input. Representative examples of such projected images for both top and QCD jet classes are shown in Fig.~\ref{fig:pca_jetImg}.

\section{Results} \label{secIII}

The results presented here are based on a qualitative level comparison. We obtain accuracy and loss results for different setups using the default.qubit simulator from PennyLane for quantum computing calculations, and the TensorFlow library for CNNs. Results are obtained separately for $SO(4)$ and $SU(4)$ quantum circuits, considering the three loss functions, Hinge, MSE, and Cross-entropy. For each loss function, in the first step, we change the encodings (TPE, HEE1, HEE2, and CHE) while keeping the batch-size fixed to 32. Then we choose the opposite, namely, we fix the encoding to HEE1 and change the batch-sizes (16, 32, 64, and 128).  

To ensure a fair comparison between QCNNs and CNNs, we aim to make their overall setups as similar as possible. The number of trainable parameters plays a significant role in achieving good classical results. Therefore, for an $SO(4)$ QCNN with \textbf{30} parameters, the corresponding number of CNN parameters is~\textbf{33}. Similarly, for an $SU(4)$ QCNN with~\textbf{48} parameters, the corresponding number for CNN is~\textbf{51}. For every run, we initialize a random set of parameters that are optimized after each epoch with a learning rate of $0.001$.

On top of the loss values, additional metric is used to measure the accuracy, which is given by the total number of correct classifications over the total number of all classifications,
that is 

\begin{equation}
    \text{(accuracy)} := \frac{\text{(correct classifications)}}{\text{(the number of total samples)}}\,.
\end{equation}

\subsection{Dependence on Encodings} \label{3.1}

Accuracy results after $30$ epochs averaged over $50$ runs for QCNN with $SO(4)$ and $SU(4)$ with varying (quantum) encoding types compared to CNN for the three loss functions are presented in Table \ref{tab1}. Here, the batch-size is fixed to 32. We can notice that accuracy results for QCNN are better than those of CNN when we use $SO(4)$ along with Hinge (except for CHE encoding), MSE, or cross-entropy. The performance of QCNN for $SU(4)$ setups is similar or worse.

\begin{table}[ht]
    \centering
    \begin{tabular}{|c|c|c|c|c|}
        \hline
        Circuit & Loss & Encoding & QCNN Acc ($\%$) & CNN Acc ($\%$) \\ \cline{1-5}
        \hline
        \multirow{12}{*}{$SO(4)$} & \multirow{4}{*}{H} & TPE & 99.13 $\pm$ 0.17 & \multirow{4}{*}{ \small{95.47  ± 1.42} } \\ \cline{3-4}
                               &                            & HEE1 & \textbf{99.84 $\pm$ 0.07} &  \\ \cline{3-4}
                               &                            & HEE2 & 97.47 ± 0.38 &  \\ \cline{3-4}
                               &                            & CHE & 94.58 ± 0.34 &  \\ \cline{2-5}
                               & \multirow{4}{*}{M} & TPE & 99.45 ± 0.11 & \multirow{4}{*}{ \small{96.26 ± 1.09} } \\ \cline{3-4}
                               &                            & HEE1 & \textbf{99.87 ± 0.06} & \\ \cline{3-4}
                               &                            & HEE2 & 98.68 ± 0.24 &  \\ \cline{3-4}
                               &                            & CHE & 96.82 ± 0.21 &  \\ \cline{2-5}
                               & \multirow{4}{*}{C} & TPE & 99.26 ± 0.14 & \multirow{4}{*}{ \small{94.16 ± 1.68}} \\ \cline{3-4}
                               &                            & HEE1 & \textbf{99.88 ± 0.06} &  \\ \cline{3-4}
                               &                            & HEE2 & 97.75 ± 0.36 &  \\ \cline{3-4}
                               &                            & CHE & 96.01 ± 0.27 &  \\ \cline{1-5}
        \multirow{12}{*}{$SU(4)$} & \multirow{4}{*}{H} & TPE & 99.68 ± 0.12 & \multirow{4}{*}{ \small{97.38 ± 0.45}} \\ \cline{3-4}
                               &                            & HEE1 & \textbf{99.88 ± 0.09} &  \\ \cline{3-4}
                               &                            & HEE2 & 99.24 ± 0.13 &  \\ \cline{3-4}
                               &                            & CHE & 96.89 ± 0.20 & \\ \cline{2-5}
                               & \multirow{4}{*}{M} & TPE & \textbf{99.91 ± 0.05} & \multirow{4}{*}{ \small{98.70 ± 0.23}} \\ \cline{3-4}
                               &                            & HEE1 & \textbf{99.94 ± 0.05} &  \\ \cline{3-4}
                               &                            & HEE2 & 99.69 ± 0.05 &  \\ \cline{3-4}
                               &                            & CHE & 97.91 ± 0.13 &  \\ \cline{2-5}
                               & \multirow{4}{*}{C} & TPE & 99.89 ± 0.07 & \multirow{4}{*}{ \small{98.12 ± 0.38}} \\ \cline{3-4}
                               &                            & HEE1 & \textbf{99.90 ± 0.07} &  \\ \cline{3-4}
                               &                            & HEE2 & 99.44 ± 0.10 &  \\ \cline{3-4}
                               &                            & CHE & 97.48 ± 0.16 &  \\ \hline
    \end{tabular}
    \caption{Test accuracy for Hinge (H), MSE (M), and Cross-entropy (C) loss function with the different encodings results averaged over 50 runs for $SO(4)$ with 30 parameters and CNN results with 33 parameters, and $SU(4)$ with 48 parameters and the comparable CNN with 51 parameters. The batch-size is fixed to 32.}
    \label{tab1}
\end{table}

We can also observe the relatively large error for CNN results at the low parameter regime. For QCNN setups, first with relatively high accuracy, HEE1 encoding (values given in bold in Table \ref{tab1}), and then TPE showed better results compared to HEE2 or CHE. 
This could be possibly understood from dataset induced barren plateau investigated in~\cite{2021arXiv211014753T}. In other words, the HEE2 and CHE encoding would be too complex specifically for this task, which induces trainability issues (barren plateau). However, we have to study performance by increasing the number of qubits to make this connection more concrete, which we leave for the future direction.

\begin{figure*}
    \centering
    \begin{subfigure}[b]{0.49\textwidth}
    \includegraphics[width=\textwidth]{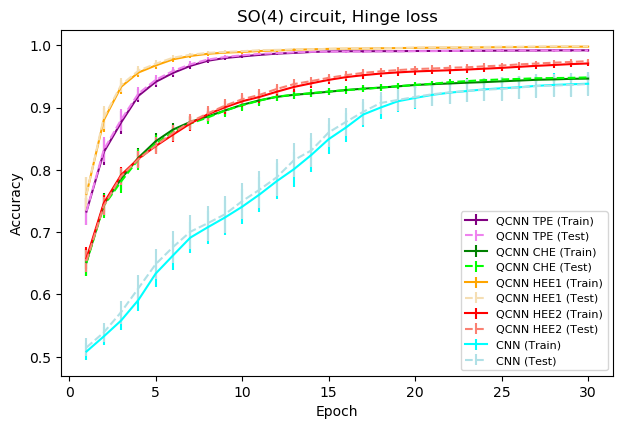}
    \subcaption{}
    \label{FigEncAccA}
    \end{subfigure}
    \centering
    \begin{subfigure}[b]{0.49\textwidth}
    \includegraphics[width=\textwidth]{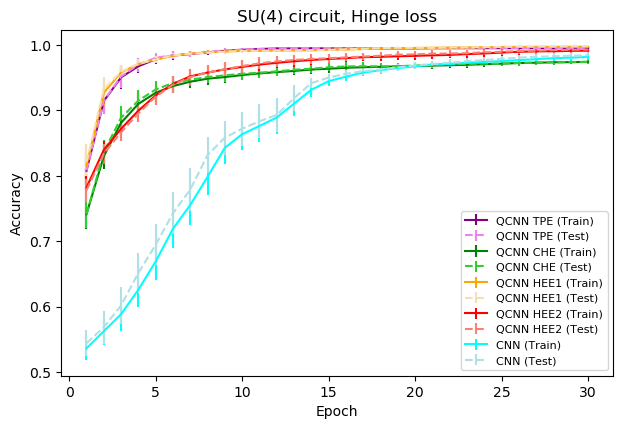}
    \subcaption{}
    \label{FigEncAccB}
    \end{subfigure}
    \centering
    \begin{subfigure}[b]{0.49\textwidth}
    \includegraphics[width=\textwidth]{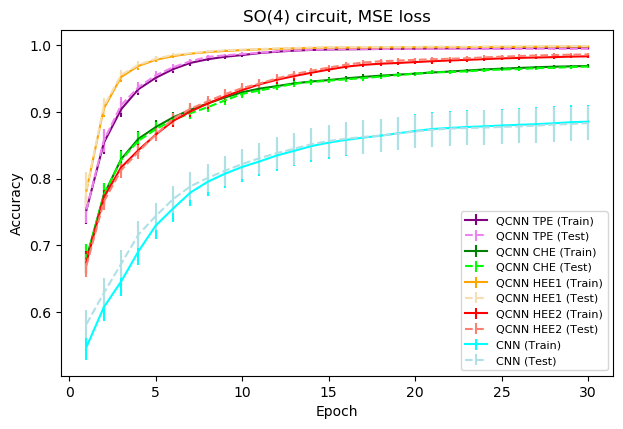}
    \subcaption{}
    \label{FigEncAccC}
    \end{subfigure}
    \centering
    \begin{subfigure}[b]{0.49\textwidth}
    \includegraphics[width=\textwidth]{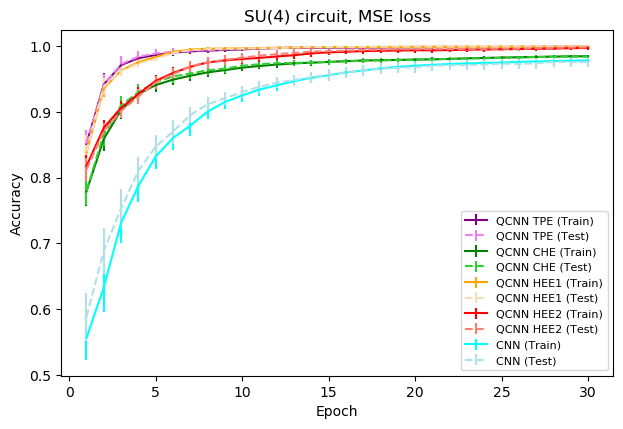}
    \subcaption{}
    \label{FigEncAccD}
    \end{subfigure}
    \centering
    \begin{subfigure}[b]{0.49\textwidth}
    \includegraphics[width=\textwidth]{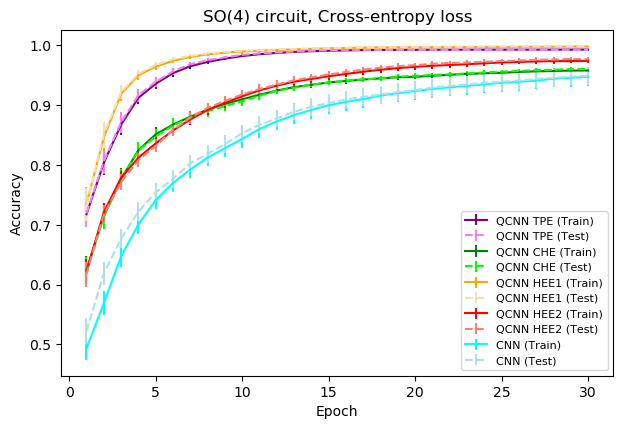}
    \subcaption{}
    \label{FigEncAccE}
    \end{subfigure}
    \centering
    \begin{subfigure}[b]{0.49\textwidth}
    \includegraphics[width=\textwidth]{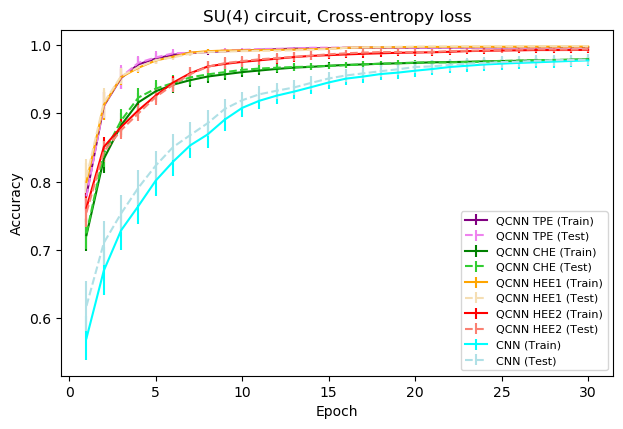}
    \subcaption{}
    \label{FigEncAccF}
    \end{subfigure}
    \caption{$SO(4)$ and $SU(4)$ QCNN accuracy encoding results compared to CNN for Hinge (a) and (b), MSE (c) and (d), and Cross-entropy (e) and (f).} 
    \label{FigEncAcc}
\end{figure*}

The accuracy curves corresponding to these settings are also summarized in Fig.~\ref{FigEncAcc} for~$SO(4)$ and $SU(4)$ compared to CNN with Hinge (Fig.~\ref{FigEncAccA} and~\ref{FigEncAccB}), MSE (Fig.~\ref{FigEncAccC} and~\ref{FigEncAccD}), and Cross-entropy (Fig.~\ref{FigEncAccE} and~\ref{FigEncAccF}) loss functions. We can notice the faster convergence at early epochs for all setups of QCNN for both $SO(4)$ and $SU(4)$ compared to CNN.

\subsection{Dependence on Batch-sizes} \label{3.2}

\begin{table}[ht!]
    \centering
    \begin{tabular}{|c|c|c|c|c|}
        \hline
        Circuit & Loss & Batch-size & QCNN Acc ($\%$) & CNN Acc ($\%$) \\ \cline{1-5}
        \hline
        \multirow{12}{*}{$SO(4)$} & \multirow{4}{*}{H} & 16 & \textbf{99.93 ± 0.07} & 94.70 ± 1.74 \\ \cline{3-5}
                               &                            & 32 &  \textbf{99.73 ± 0.12} & 92.48 ± 1.79 \\ \cline{3-5}
                               &                            & 64 & 99.23 ± 0.23 & 89.94 ± 2.02 \\ \cline{3-5}
                               &                            & 128 & 97.48 ± 0.82 & 69.30 ± 2.90 \\ \cline{2-5}
                               & \multirow{4}{*}{M} & 16 & \textbf{100.00 ± 0.00} & 94.12 ± 2.00\\ \cline{3-5}
                               &                            & 32 & \textbf{99.94 ± 0.03} & 94.76 ± 1.84 \\ \cline{3-5}
                               &                            & 64 & 99.69 ± 0.10 & 87.38 ± 2.48 \\ \cline{3-5}
                               &                            & 128 & 98.89 ± 0.34 & 87.35 ± 1.33\\ \cline{2-5}
                               & \multirow{4}{*}{C} & 16 & \textbf{100.00 ± 0.00} & 92.48 ± 2.30 \\ \cline{3-5}
                               &                            & 32 & \textbf{99.91 ± 0.05} & 92.93 ± 1.92 \\ \cline{3-5}
                               &                            & 64 & 99.43 ± 0.16 & 88.54 ± 2.47 \\ \cline{3-5}
                               &                            & 128 & 98.23 ± 0.54 & 83.38 ± 1.97 \\ \cline{1-5}
        \multirow{12}{*}{$SU(4)$} & \multirow{4}{*}{H} & 16 & \textbf{99.75 ± 0.14} & 97.95 ± 0.44 \\ \cline{3-5}
                               &                            & 32 & \textbf{99.92 ± 0.08} & 97.56 ± 0.45 \\ \cline{3-5}
                               &                            & 64 & 99.72 ± 0.11 & 93.36 ± 1.63 \\ \cline{3-5}
                               &                            & 128 & 99.22 ± 0.17 & 85.35 ± 2.06 \\ \cline{2-5}
                               & \multirow{4}{*}{M} & 16 & \textbf{99.84 ± 0.09} & 98.80 ± 0.29 \\ \cline{3-5}
                               &                            & 32 & \textbf{99.94 ± 0.05} & 98.32 ± 0.37 \\ \cline{3-5}
                               &                            & 64 & 99.86 ± 0.06 & 97.02 ± 0.46 \\ \cline{3-5}
                               &                            & 128 & 99.64 ± 0.09 & 90.88 ± 1.34 \\ \cline{2-5}
                               & \multirow{4}{*}{C} & 16 & \textbf{99.82 ± 0.10} & 97.16 ± 1.13 \\ \cline{3-5}
                               &                            & 32 & \textbf{99.90 ± 0.07} & 96.63 ± 1.26 \\ \cline{3-5}
                               &                            & 64 & 99.79 ± 0.08 & 95.86 ± 0.68 \\ \cline{3-5}
                               &                            & 128 & 99.35 ± 0.14 & 91.57 ± 0.89 \\ \hline
    \end{tabular}
    \caption{Test accuracy batch-sizes results averaged over 50 runs for $SO(4)$ with 30 parameters and their corresponding CNN with 33 parameters, and $SU(4)$ with 48 parameters and the comparable CNN with 51 parameters. Encoding is fixed to HEE1.}
    \label{tab2}
\end{table}

\begin{figure}[htbp!]
    \centering
    \includegraphics[width=0.48\textwidth]{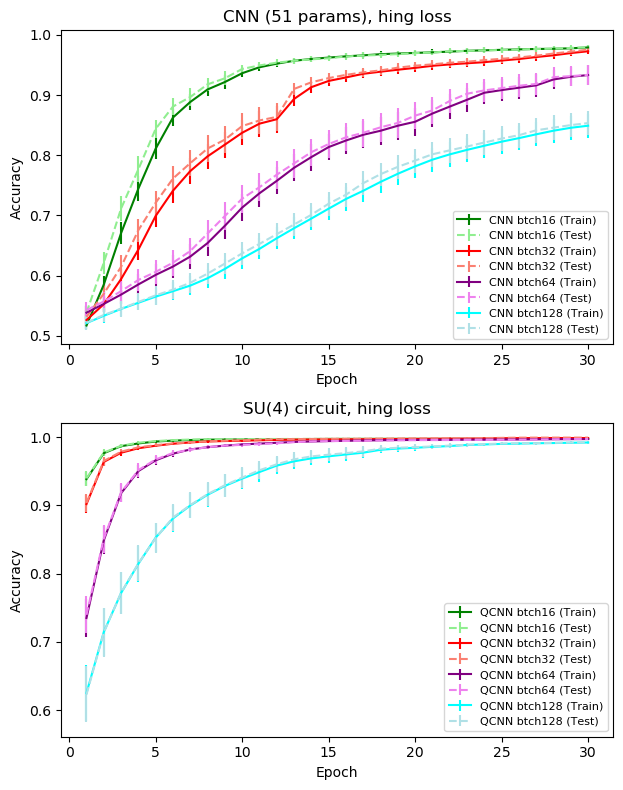}
    \caption{Accuracy results of the different batch-sizes in CNN (top) and $SU(4)$ QCNN (bottom) for Hinge loss.} 
    \label{FigBtchAcc}
\end{figure}

After fixing the encoding to HEE1 which shows the best performance for $32$ batch-size, we varied the batch size to four different sizes. The results averaged over $50$ runs are displayed in Table~\ref{tab2}. Again we observe a tendency for QCNN to have a better performance for all the batch-sizes in $SO(4)$ compared to CNN, and results are comparable for $SU(4)$ and CNN. However, what is clearly noticeable in most of the cases, is that as the batch size increases, the accuracy tends to decrease. This observation is also depicted in Fig. \ref{FigBtchAcc} for $SU(4)$ and the corresponding CNN model with~$51$ parameters for Hinge loss as an example. We also observe that a batch size of~$16$ has faster convergence compared to the other batch sizes, while a batch size of~$128$ exhibits the slowest convergence. Batch sizes of 16 and 32 demonstrate similar performance (values shown in bold in Table \ref{tab2}), which is superior to both 64 and 128. It is essential to note that slower or faster convergence does not necessarily equate to slower or faster training times, since the time it takes to train each epoch is different from quantum to classical.

\subsection{DEA circuit}

\begin{figure}[ht!]
    \centering
    \includegraphics[width=0.45\textwidth]{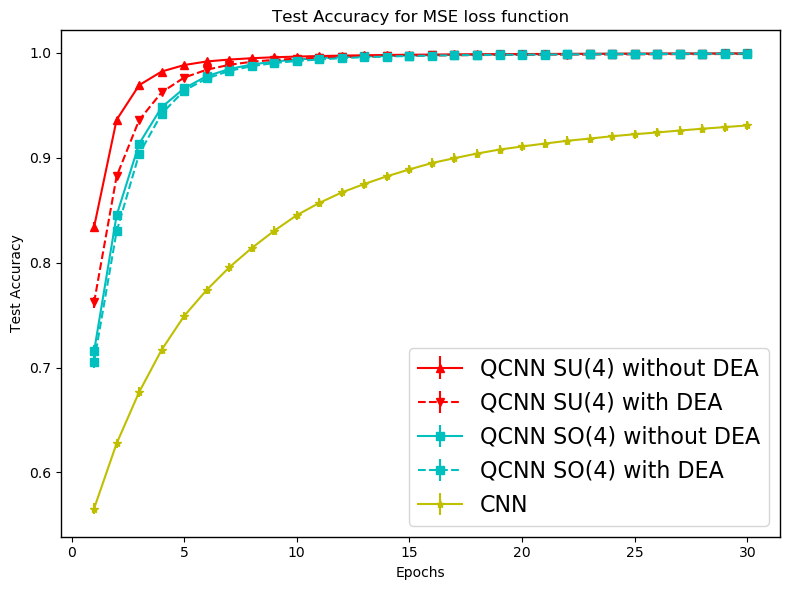}
    \caption{Test accuracy results of the different circuits including and excluding DEA compared to CNN.} 
    \label{FigDEA}
\end{figure}

QCNN with the $SU(4)$ convolutional circuit demonstrated better performance compared to the $SO(4)$ convolutional circuit. However, it required more parameters and, hence, longer training times. Now the question is, will we get similar results if we train the QCNN $SU(4)$ circuit after removing the redundant parameters using DEA?

For this task, we select the setup that resulted in the best accuracies for both QCNNs and CNNs. This setup, as shown in Table \ref{tab1}, corresponds to the use of MSE loss along with the HEE1 encoding. After performing a DEA check on the $SU(4)$ QCNN circuit with originally 48 parameters, 17 of them were found to be redundant, leaving a total of 31 parameters in the optimized circuit. This is compatible with the dimension of the state space ($2^{Q+1}-1$, where $Q$ is the number of qubits in the system). Therefore, the circuit is minimal in the set of trainable parameters and maximally expressive.

Test accuracy results of the MSE loss averaged over 1000 runs for $SU(4)$ and $SO(4)$ QCNN circuits before and after applying DEA, compared to CNN accuracy can be found in Table \ref{tab3} and Fig.~\ref{FigDEA}. The accuracy values at epoch 30 in Table~\ref{tab3} indicate a clear difference between the case when we have comparable number of parameters (31) of the DEA circuit with the CNN model (33 parameters) where the DEA circuit has higher accuracy value. The curves in Fig.~\ref{FigDEA} also depict this result between all QCNN circuits (with and without DEA) compared to the CNN accuracy. We may notice slight differences between the $SU(4)$ circuit with and without DEA and the $SO(4)$ circuit at early epochs. The $SU(4)$ circuit without DEA has more parameters (48) which help it converge to the solution after fewer training epochs compared to the case with less parameters in the circuit with DEA (31). Both $SO(4)$ circuits, with and without DEA have a similar performance, and again we notice the slightly higher accuracy at early epochs for the circuit with the higher number of parameters. However, the $SU(4)$ circuit with DEA, has much less parameters compared to its original circuit and yet it shows comparable accuracy values. The $SO(4)$ circuit has 30 parameters, which is very close to the 31 parameters in the $SU(4)$ circuit with DEA. Notably, the DEA circuit shows comparable performances and even improved accuracy in the early epochs, suggesting that DEA can reduce the number of trainable parameters while maintaining strong performance.

\begin{table}[ht!]
    \centering
    \begin{tabular}{|c|c|c|}
        \hline
        Circuit structure & params. & Acc ($\%$) \\ \cline{1-3}
        \hline
          $SU(4)$ without DEA    & 48 & 99.95 ± 0.01 \\ \cline{1-3}
          $SU(4)$ with DEA   & 31 & 99.89 $\pm$ 0.01  \\  \cline{1-3} 
          $SO(4)$ without DEA    & 30 & 99.95 ± 0.01 \\ \cline{1-3}
          $SO(4)$ with DEA  & 25 & 99.93 ± 0.01  \\ \cline{1-3}
          CNN           & 33 & 93.08 ± 0.44  \\ \cline{1-3}
    \end{tabular}
    \caption{Test accuracy results at epoch 30 averaged over 1000 runs using MSE loss function and HEE1 embedding with the various circuits studied.}
    \label{tab3}
\end{table}

\subsection{Effect of PCA and model complexity}

One of the main advantages of PCA is its ability to preserve the variance in the dataset after dimensionality reduction. The variance preserved by PCA can be quantified by calculating the cumulative explained variance, which can be used as a criterion for selecting the number of principal components. Since the variance of each principal component is given by its corresponding eigenvalue ($\lambda_k$), the amount of variation explained by the chosen first (m) components is calculated as the sum of their eigenvalues divided by the sum of all eigenvalues~\cite{jolliffe2016principal}
\[
\text{Explained Variance (\%)} =
(\displaystyle\sum_{k=1}^{m}\lambda_k 
\bigg/
\displaystyle\sum_{k=1}^{p}\lambda_k)
\times 100, 
\]
where $p$ is the total number of principal components. In the previous results of using 4 PCA components, the preserved variance of the dataset was 50.54\%. Despite this substantial reduction in dimensionality, the classification performance remained robust. Increasing the number of PCA components to 16, for example, increases the preserved variance to 68.67\%. However, increasing the number of components does not necessarily improve the accuracy of the QCNN model. Following our predefined QML setup, using 16 components would require a quantum circuit with 16 qubits. A detailed discussion of how the results change depending on the number of PCA components, and consequently the number of qubits, can be found in~\cite{elhag}. In the following, we investigate how the complexity of the classical model affects the classification accuracy. This is done by applying a Multilayer Perceptron (MLP)~\cite{MLP} to the PCA components before passing the resulting input features to the CNN model.

\begin{figure}[th!]
    \centering
    \includegraphics[width=0.45\textwidth]{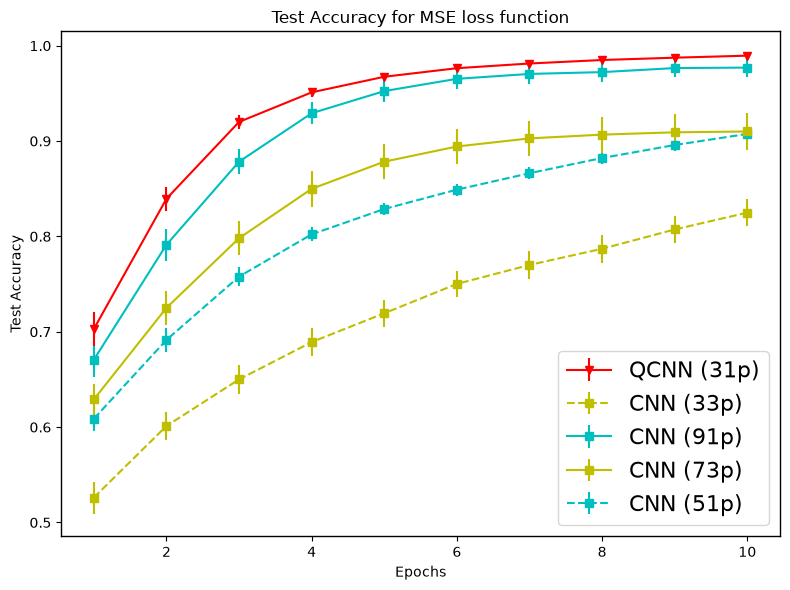}
    \caption{Test accuracy results of the QCNN with 31 parameters compared to CNN models with 33, 51, 73, and 91 parameters.} 
    \label{FigComplexCNN}
\end{figure}

Combining the same CNN models presented previously (see Fig.~\ref{fig:CNN_struc}) with an MLP increases the model complexity and introduces additional trainable parameters. In this case, the MLP is implemented as two additional dense layers applied to the PCA features before passing their output to the CNN model. The overall number of trainable parameters in the CNN models with 33 and 51 parameters consequently increases to 73 and 91, respectively. Training these models on the same dataset produces the test accuracy results, averaged over 100 runs, presented in Fig.~\ref{FigComplexCNN}. The results clearly show an improvement in the accuracy of the CNN model with 73 parameters compared to the 33-parameter and the 91 compared to the 51-parameter model. However, more notably, the DEA-optimized QCNN with only 31 parameters still outperforms the 73-parameter CNN model and achieves comparable performance to the model with 91 parameters. Although the plot indicates that increasing the number of parameters and, consequently, the complexity of the CNN model can improve its accuracy, the results also demonstrate that a lower parameter count can achieve comparable classification performance.

\section{Conclusion and outlook} \label{secIV}

In this paper, we utilized two QCNN architectures one with $SO(4)$ convolutional circuit and the other built with an $SU(4)$ convolutional circuit.
We varied the setups of each architecture to test different loss functions, encodings, and batch-sizes. We compared each setup with a corresponding CNN model. 
From the preliminary results we got for our quantum and classical models, we found that the QCNN, in the absence of noise and using a classical simulator, performs better than the corresponding CNN model for most choices of data encoding, batch-size, loss function, and convolutional block involved. We observed that increasing the complexity of the classical model and the number of trainable parameters improves its accuracy.
However, it is important to emphasize that the system we studied here is still relatively simple, both in QCNN and CNN. As a result, it was possible to train the classical model with a very small number of parameters, whereas the usual case of training CNN models is to use a high number of parameters that may reach up to millions.
Besides, we only performed the noiseless simulation and the effects from noise would spoil the QCNN performance unless we can tame them.

In our study, DEA demonstrated the ability to significantly reduce the number of trainable parameters while maintaining good overall accuracy, thus enabling the QCNN to achieve higher accuracy compared to the CNN model with a similar parameter count. This could have a great impact when one is aiming for noisy hardware implementation, in which it will reduce the complexity of a circuit, and hence the computational complexity on the real device. Another way of reducing the complexity of the problem is through equivariant QNN (EQNN)~\cite{Forestano_2024} which mainly relies on the symmetry found in data. EQNN methods allow us to construct QNNs that automatically satisfy the desired symmetry constraints. Since our dataset can also have rotational symmetry thus potentially, we can apply these techniques.

It is worth noting that, despite the improved accuracy we found for QCNN over the CNN model, we are not claiming that our model outperforms all the existing models that did not achieve comparable accuracies for the top-tagging task in general. This is because the accuracy results depend heavily on the specific dataset used and the details of its preprocessing. Hence, a direct, one-to-one comparison is not possible unless fair training conditions with the same dataset were performed. For instance, combining multiple jets from a single class into a single image results in a characteristic structure for that class. However, in real HEP experiments, one would need to classify single jets produced from an event, which is a more challenging task. That said, we believe our model holds promising potential for jet image classification tasks in a more general manner. In particular, if the dataset was pre-processed such that when PCA is applied, the principal components can still capture most of the important features existed in the image. 

It is also important to note that, since the results presented in this work are obtained using noiseless classical simulation, the effects of noise present in real quantum hardware, finite-shot sampling, circuit depth, and other hardware effects are not considered. Although the results demonstrate that the proposed QCNN can achieve competitive classification accuracy with a reduced number of trainable parameters, this should not be interpreted as demonstrating a practical quantum advantage. The results should instead be understood as an evaluation of the performance of the proposed quantum model under ideal simulation conditions. Investigating the effect of realistic hardware noise and implementation constraints would provide a more practical assessment of the proposed approach.

As QCNN has been indicated to be classically simulable~\cite{bermejo2024quantumconvolutionalneuralnetworks} for the classification of classical data, and at the same time it showed improved classification accuracy over the conventional classical CNN models for a number of case studies, one could use this as an advantage to study the scaling of performance and resources with respect to the complexity of the dataset and/or the number of qubits. The classical simulability helps in avoiding using real quantum hardware, which makes it possible to use a much higher number of qubits via classical simulation, and hence, for our problem of jet images classification, we can use higher dimensional images without going through PCA which is a more natural practice of dealing directly with pixels of the original dataset. 

That said, other studies such as \cite{xiao} argues that QML can still demonstrate a practical quantum advantage. In their study, they used physically generated 1-D structured data. Their quantum model achieved better task performance and parameter efficiency than comparable classical models, rather than simply demonstrating a theoretical quantum advantage.

Another interesting direction would be to apply QCNN and investigate other QML models on the use of quantum dataset which could also help in analysing experimental HEP problems such as studying the quantum properties of the final state particles detected from a collision of interest e.g. spin entanglement \cite{2024}. In support to this, a work by \cite{Xu} illustrated the potential of QML for inherently quantum data by using quantum reinforcement learning to process quantum many-body states, enabling improved quantum-sensing performance without the critical slowing down associated with conventional adiabatic methods. 

\section*{Code availability}

The code used for data preprocessing, model implementation, and training, which produced the results reported in this work, is available at \url{https://github.com/haelhag/QCNNs4JetImagesClassification.git}. The repository includes scripts and documentation to allow reproduction of the results and further investigation.

\begin{acknowledgments}
This work is supported with funds from the Ministry of Science, Research, and Culture of the State of Brandenburg within the Centre for Quantum Technologies and Applications (CQTA). 
This work is also supported by the Center of Innovations for Sustainable Quantum AI (JST Grant Number JPMJPF2221).
\begin{center}
    \includegraphics[width = 0.08\textwidth]{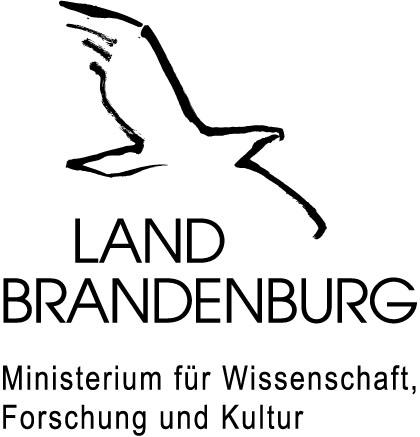}
\end{center}

ADT gratefully acknowledges the support of the Physikalisch-Technische Bundesanstalt.
\end{acknowledgments}

\nocite{*}

\clearpage
 
\bibliographystyle{apsrev4-2}
\bibliography{apssamp}

\end{document}